\newif\ifeptcs
\eptcstrue

\ifeptcs
 \documentclass[submission, copyright, creativecommons]{eptcs}
\else
\documentclass{article}
\fi

\usepackage[english]{babel}
\ifeptcs
\else
\usepackage[blocks]{authblk}
\fi

\ifeptcs
\else
\usepackage[letterpaper,top=2cm,bottom=2cm,left=2.5cm,right=2.5cm,marginparwidth=1.75cm]{geometry}
\fi

\usepackage{amsmath}
\usepackage{graphicx}

\ifeptcs\else
\usepackage[colorlinks=true, allcolors=blue]{hyperref}
\fi

\usepackage{amsmath}
\usepackage{algorithm}
\usepackage[noend]{algpseudocode}
\usepackage{enumitem}
\usepackage{graphicx}
\usepackage{subcaption}

\makeatletter
\def\BState{\State\hskip-\ALG@thistlm}

\makeatother
\ifeptcs

\title{Modelling and Search-Based Testing of Robot Controllers Using Enzymatic Numerical P Systems}

\author{Radu Traian Bobe \quad\quad Florentin Ipate \quad\quad Ionu\c{t} Mihai Niculescu \institute{{Department of Computer Science, Faculty of Mathematics and Computer Science}, {University of Bucharest}, \\ {Str Academiei 14}, {Bucharest}, {010014}, {Romania}}\email{radu.bobe@s.unibuc.ro florentin.ipate@unibuc.ro ionutmihainiculescu@gmail.com}}
\def\titlerunning{Modelling and search-based testing of robot controllers using enzymatic numerical P systems}
\def\authorrunning{Radu Traian Bobe, Florentin Ipate \& Ionu\c{t} Mihai Niculescu}

\else

\author{Radu Traian Bobe}

\author{Florentin Ipate}

\author{Ionu\c{t} Mihai Niculescu}

\affil{{Department of Computer Science, Faculty of Mathematics and Computer Science, University of Bucharest}, {Str Academiei 14}, {Bucharest}, {010014}, {Romania}}

\affil{radu.bobe@s.unibuc.ro, florentin.ipate@unibuc.ro, \\ ionutmihainiculescu@gmail.com}
\fi

\date{}

\begin{document}
\maketitle

\begin{abstract}
The safety of the systems controlled by software is a very important area in a digitalized society, as the number of automated processes is increasing. In this paper, we present the results of testing the accuracy of different lane keeping controllers for an educational robot. In our approach, the robot is controlled using numerical P systems and enzymatic numerical P systems. For tests generation, we used an open-source tool implementing a search-based software testing approach.
\end{abstract}


\section{Introduction}
Due to the remarkable technological progress of late years, software applications tend to have a considerable role in solving most problems of everyday life. The medical, financial or automotive fields are just three of the main areas in which software products are intensively used. Given the importance of these areas in every individual's life, ensuring product quality and functionality is an essential step in the development process. The safety of software systems for large-scale use is ensured by testing. Software testing aims to validate the fulfillment of the requirements defined for the developed product, as well as to identify possible unwanted behaviors triggered by simulating certain operational contexts.\\

In this paper, we propose an approach for testing two different lane keeping controllers designed to move an educational robot called \textit{E-puck} \cite{mondada2009puck}. Both controllers are based on numerical P systems, introduced by G. Păun and R. Păun in \cite{puaun2006membrane}. We also provide an equivalent version for the models using enzymatic numerical P systems, an extension of numerical P systems, defined by A. Pavel et al. in \cite{pavel2010enzymatic}. For this experiment we used some reliable tools which will be introduced in the following sections. 

The paper is structured as described: Section 2 presents the P system variants to be used in the paper. Section 3 introduces the working environment, including the tools used. Section 4 describes the models and the main differences between them, while Section 5 illustrates the testing approach along with the results. In the end, Section 6 presents the future work and conclusions.

\section{Preliminaries}
\textit{Membrane computing} is a field of research introduced by Gh. Păun in \cite{paun2002membrane, paun2010membrane}. The computational paradigm was originally inspired by the structure and functionality of the living cells. Several classes of membrane systems (P systems) have been later defined and investigated, being classified according to the structure of the membranes as \textit{cell-like}, \textit{tissue-like} and \textit{neural-like} P systems. Membrane computing has made significant breakthroughs in the last decades in fields like computer science, economics or biology. Depending on the requirements, extensions of the main concept were introduced and our experiment involves two of these types: numerical P systems and enzymatic numerical P systems.

\subsection{(Enzymatic) Numerical P System}

The (enzymatic) numerical P systems \cite{puaun2006membrane, pavel2010enzymatic} are computational models that only inherit the membrane structure from the membrane systems, more exactly a cell-like membrane structure. The membranes contain \textit{variables} and their values are processed by the \textit{programs} on every time unit. The whole system is synchronized by a global clock in discrete time units.\\

The (enzymatic) numerical P system (EN P system) is defined by the tuple:

\begin{equation}\label{nps}
  \varPi \ \mathit{= (m, H, \mu, (Var_1, Pr_1, Var_1(0)),\dots,(Var_m, Pr_m, Var_m(0)))}
\end{equation}

where:
\begin{itemize}
    \item $\mathit{m} \geq 1$ is degree of the system $\Pi$ (the number of membranes);
    \item \textit{$\mathit{H}$} is an alphabet of labels;
    \item $\mu$ is membrane structure;
    \item $\mathit{Var_i}$ is a set of variables from membrane $\mathit{i, 1 \le i \le m}$;
    \item $\mathit{Var_i(0)}$ is the initial values of the variables from region $\mathit{i, 1 \le i \le m}$;
    \item $\mathit{Pr_i}$ is the set of programs from membrane $\mathit{i, 1 \le i \le m}$. \\ \\
    The program $\mathit{Pr_{l_i,i}}$, $1 \leq \mathit{l_i} \leq \mathit{m_i}$ has one of the two following forms:
    \begin{enumerate}[label=\roman*)]
        \item non-enzymatic \\
        $$\mathit{F_{l_i,i}(x_{1,i},\dots,x_{k,i}) \rightarrow c_{1,i}|v_1+c_{2,i}|v_2+ \dots + c_{m_i,i}|v_{m_i}}$$ \\
        where $\mathit{F_{l_i,i}(x_{1,i},\dots,x_{k,i})}$ is the production function, $\mathit{c_{1,i}|v_1+c_{2,i}|v_2+ \dots + c_{m_i,i}|v_{m_i}}$ is the repartition protocol, and $\mathit{x_{1,i},\dots,x_{k,i}}$ are variables from $Var_i$. Variables $v_1, v_2 \dots v_{m_i}$ can be from the region where the programs are located, and to its upper and inner compartments, for a particular region $i$. If a compartment contains more than one program, only one will be chosen in non-deterministically manner.
        \item enzymatic \\
        $$\mathit{F_{l_i,i}(x_{1,i},\dots,x_{k,i})|_{e_i} \rightarrow c_{1,i}|v_1+c_{2,i}|v_2+ \dots + c_{m_i,i}|v_{m_i}}$$ \\
        where $\mathit{e_i}$ is an enzymatic variable from $\mathit{Var_i}$, $\mathit{e_i \notin \{x_{1,i},\dots,x_{k,i}, v_1,\dots, v_{m_i}\}}$. The program can be applied at time $\mathit{t}$ only if $\mathit{e_i > min(x_{1,i}(t),\dots,x_{k,i}(t))}$. The programs that meet this condition in a region will be applied in parallel.
    \end{enumerate}
\end{itemize}

When the program is applied by the system at time $t \geq 0$, the computed value

$$ \mathit{q_{l_i,i}(t) = \frac{F_{l_i,i}(x_{1,i}(t),\dots,x_{k,i}(t))}{ \displaystyle\sum_{j=1}^{n_i} c_{j,i} }} $$

represents the \textit{unitary portion} that will be distributed to the variables $\mathit{v_1,\dots,v_n}$, proportional to coefficients $\mathit{c_{1,i},\dots,c_{m_i,i}}$, where $\mathit{c_{j,i} \in \mathbf{N}^+}$ and the received values will be $\mathit{q_{l_i,i}(t) \cdot c_{1,i},\dots, q_{l_i,i}(t) \cdot c_{m_i,i}}$.


The values of variables, from $\mathit{t - 1}$, present in the production functions are \textit{consumed}, reset to zero, and their new value is the sum of the proportions distributed to variable through the repartition protocols, if they appear in them or remain at the value zero.

\section{Experimental environment}

In this section we will provide brief descriptions of the tools we integrated in our experiment. Firstly, we used an open-source software which allows the simulation of numerical P systems and enzymatic numerical P systems. The simulator is called PeP and will be introduced later in this section.  Since we don't have the physical education robot involved in this study, we also used a dedicated platform for robot simulations, called Webots. For tests generation, we used a tool which won the \textit{SBST Tool Competition 2022} \cite{gambi2022sbst}. We will discuss later in this section the arguments for using a search-based testing tool. 

\subsection{PeP simulator}
 PeP simulator \cite{pep} is an open-source product developed by A.Florea and C.Buiu,   used for simulations based on numerical P systems and enzymatic numerical P systems. The program is written in Python and receives numerical P systems as an input file. The input file includes the membrane structure and the contents of each membrane, being stored in memory and executed. \\
 PeP can be used as a stand-alone tool for simple simulations and run from the command line with some options, like the number of simulations steps or a csv document generation containing the values at each step of the simulation. As observed in \cite{pep}, the tool comes with a set of basic input files examples, both numerical P systems and enzymatic numerical P systems.  \\
 
 Besides the simplicity of running this tool, another advantage which can be taken into account when using PeP is that it can be used as an integrated module in complex projects. We used this approach in our experiment in order to make a controller accepted by the robot simulation platform and able to receive information from the platform. In our lane keeping experiments, the simulation ends when the robot drives off the generated lane or when the lane is kept until the end. 
 
 \subsection{Webots and E-puck}

 Webots is a robotics simulation software which allows the user to construct a complex environment for programming, modelling and simulating mobile robots. The environment can include multiple scene objects with different properties which can be set from the graphic interface or from the generation files \cite{michel2004cyberbotics}. In addition, the robots can be equiped with a large number of objects called nodes, like sensors, camera, GPS, LED, light sensor etc.\\
 In our approach, additionally to the original equipments of E-puck, we used a GPS attached to the turret slot in order to examine the coordinates at each step of the simulation. The scenes, called "worlds" in Webots, are containing the road that the robot will try to follow. The roads are generated with Ambiegen, a tool that will be described later in this section. Each world is defined by a \emph{.wbt} file. The objects can be edited in this file and we used this option in order to place the road object in the scene with coordinates exported from Ambiegen. Additional functionalities, like sensors or GPS can also be added to the robot by editing the world file which will be imported in Webots, obtaining the visual representation of the scene.
 
As mentioned before, for this experiment we used a robot widely known from educational and research purposes, called E-puck. At the moment, the robot has some capabilities that are not implemented in Webots, but considering the fact that both hardware and software components of E-puck are open source, this remains a challenging opportunity \cite{couceiro2014bridging}. 

E-puck has eight infrared proximity sensors placed around the body \cite{mondada2009puck}. For lane keeping simulation, we used just six of them: the two sensors placed in front and the four placed two on each side. This aspect can be easily adapted by changing the membrane structure and creating new membranes if more sensors are needed or deleting a few of them if required. Each sensor has a corresponding membrane in the numerical P system model and the association was made in the controller. The robot has two motors attached to the body along with two wheels, and the speed value is also changeable from the controller. 

\subsection{Ambiegen}

Ambiegen is an open-source tool that utilizes evolutionary search for the generation of test scenarios for autonomous systems. It can be used in experiments involving lane keeping assist systems and robots navigating a room with obstacles \cite{humeniuk2023ambiegen}. The software is developed in Python and uses evolutionary search \cite{whitley1996evaluating} for tests generation. 
The main goal of Ambiegen in this approach is to generate roads as test cases in order to challenge E-puck to keep the lane. The tool exports the roads in separate text files as a sequence of points, representing the road spine. From this points, we can build the road with a proportional size to E-puck. 

Challenging different LKAS (Lane Keeping Assist Systems) involves a large diversity of road topologies in order to detect the behavior in limit situations, such as narrow curves. Ambiegen figures out the solution for diversity by using a multi-objective genetic algorithm for search-based test generation, called Non-dominated Sorting Genetic Algorithm-II (NSGA-II) \cite{deb2002fast}. In Ambiegen implementation, NSGA-II has two-objectives: to increase the fault revealing power of test cases and to preserve their diversity \cite{humeniuk2022ambiegen}. This multi-objective approach, which combines roads generation with a high attention to diversity along with remarkable results at the competition mentioned above, attracted our curiosity to integrate Ambiegen with Webots and testing E-puck on the roads resulted. \\

\subsection{Experimental Procedure}

Considering the above information, we will detail the way we worked with the presented tools. PeP and Ambiegen are developed in Python and so is the robot controller. \\
First of all, we could easily integrate PeP with E-puck controller using the PeP module 
which allowed us to parse the numerical P system model as an input file for controller. Achieving this, the model membranes were associated with controller variables. Names and constant values (e.g., robot cruise speed) were taken using a text file containing membrane's values of the variables. The values were chosen empirically.  

Next is a pseudocode version of the main loop in our controller, which performs the simulation steps. 
\begin{algorithm}
 \caption{ Simulation steps performing algorithm}\label{simulation}
\begin{algorithmic}[1]

\Repeat
 \For {\text{i=1 to number\_of\_sensors}}
    \State $\textit{sensor\_membrane}(i) \gets \textit{value}(i)$
  \EndFor
 \State $\text{run one simulation step}$
 \State $\text{read } \textit{lw, rw } $ $\text{from P system} $
 \State $\textit{leftMotor} \gets \textit{lw}$
 \State $\textit{rightMotor} \gets \textit{rw}$
\Until{ the end of the road or E-puck goes out of the road} 

\end{algorithmic}
\end{algorithm}\\

Another challenge for us was to move the robot on the roads exported from Ambiegen with the above presented approach. Ambiegen exports the roads as \textit{.json} files along with informations like test outcome, maximum curvature coefficient etc. We took the road points from the file and wrote them in the world file.

Webots provides the 
possibility to extend the set of scene nodes by adding custom nodes created by users.
The mechanism is called PROTO and is described in \cite{proto}. After a node is extending with the PROTO interface, it can be instantiated from the Webots graphic interface. 

We used this technique to retrieve the points forming the spines of the roads generated by Ambiegen and putting them into the \textit{wayPoints} of the \textit{Road} node. Using javascript, used as scripting language by the PROTO, we constructed new nodes illustrating the roads from Ambiegen.  Then, in the graphic interface of Webots, the road is represented in accordance with the road from Ambiegen. With minimal Python code additions we plotted each generated road with the corresponding spine to confirm that the shape illustrated in Webots respects the original one. 

\section{Models}

In this section we will present two models used to control the robot, the core of the controller. The controller receives data from proximity sensors, that measure distances to obstacles from the environment, to determine the direction of movement of a differential two wheeled robot, E-puck, in our case.

The proximity sensor has a range of 4 cm; if the obstacles are further than this limit the sensor returns the value of 0. The proximity sensors are placed on the left and right side of the robot in the direction of its movement at different angles.

The first model was taken from \cite{buiu2017controller} and adapted. The equations that calculate the linear and angular velocity are shown below:

\begin{align*}
    \mathit{leftSpeed} =& \mathit{cruiseSpeed} + \displaystyle\sum_{i=1}^n \mathit{weightLeft_i} \cdot \mathit{prox_i} \\
    \mathit{rightSpeed} =& \mathit{cruiseSpeed} + \displaystyle\sum_{i=1}^n \mathit{weightRight_i} \cdot \mathit{prox_i}
\end{align*}

The $\mathit{leftSpeed}$ and $\mathit{rightSpeed}$ are the speeds of the two wheels of the robot. The enzymatic numerical P system described below encapsulates this behavior.\\ The first model is defined as follows:
$$\varPi  \mathit{_{M_1} = (m, H, \mu, (Var_1, Pr_1, Var_1(0)),\dots,(Var_m, Pr_m, Var_m(0)))}$$

where:

\begin{itemize}
    \item $\mathit{m = k \cdot 3 + 3, k = 6}$, where $\mathit{k}$ is the number of proximity sensors;
    \item $\mathit{H = \{s, s_c\} \cup \displaystyle\bigcup_{i=1}^{k} \{c_i, s_i, w_i\}}$;
    \item $\mathit{\mu = [ [ []_{s_1} []_{w_1}]_{c_1} \dots  [ []_{s_k} []_{w_k}]_{c_k} []_{s_c} ]_s;}$
    \item $\mathit{Var_s = \{x_{s_l}, x_{s_r}\}, Var_{s_c} = \{ x_{s_c}\},}$ \\ \hspace*{-1mm} $\mathit{ Var_{c_i} = \{x_{c_i,s_l}, x_{c_i,s_r}, x_{c_i,w_l}, x_{c_i,w_r}, e_{c_i}\}, 1 \leq i \leq k,}$ \\ \hspace*{-1mm} $ \mathit{Var_{s_i} = \{x_{s_i,i}\}, 1 \leq i \leq k,}$ \\ \hspace*{-1mm} $ \mathit{Var_{w_i} = \{x_{w_i,w_l}, x_{w_i,w_r}, e_{w_i}\}, 1 \leq i \leq }k$;
    \item $\mathit{Var_i(0) = 0, 1 \le i \le k}$;
    \item $\mathit{Pr_s = \{0 \cdot x_{s_l} \cdot x_{s_r} \rightarrow 1|x_{s_l} + 1|x_{s_r}\}}$; \\
    $\mathit{Pr_{s_c} = \{3x_{s_c} \rightarrow 1|x_{s_c} + 1|x_{s_l} + 1|x_{s_r}\}}$; \\
    $\mathit{ Pr_{c_i} = \{ x_{c_i,s_l}  \cdot x_{c_i,w_l}|_{e_{c_i}} \rightarrow 1|x_{s_l},}$\\ \hspace*{12mm} $\mathit{x_{c_i,s_r}  \cdot x_{c_i,w_r}|_{e_{c_i}} \rightarrow 1|x_{s_r} \}, 1 \leq i \leq k}$; \\
    $\mathit{Pr_{s_i} = \{3x_{s_i,i} \rightarrow 1|x_{s_i,i} + 1|x_{c_i,s_l} + 1|x_{c_i,s_r}\}, 1 \leq i \leq k}$; \\
    $\mathit{ Pr_{w_i} = \{ 2x_{w_i,w_l}|_{e_{w_i}} \rightarrow 1|x_{w_i,w_l} + 1|x_{c_i,w_l},}$\\ \hspace*{12mm} $\mathit{2x_{w_i,w_r}|_{e_{w_i}} \rightarrow 1|x_{w_i,w_r} + 1|x_{c_i,w_r} \}, 1 \leq i \leq k}$;
\end{itemize}

The meaning of the variables from the model is the following:

\begin{itemize}[label=$\circ$]
    \item $\mathit{x_{s_l}}$ and $\mathit{x_{s_r}}$ from the region $\mathit{s}$ represent $\mathit{leftSpeed}$ and $\mathit{rightSpeed}$, the sum of the products are accumulated in $\mathit{s}$ ;
    \item $\mathit{x_{s_c}}$ from the compartment $\mathit{s_c}$ is $\mathit{cruiseSpeed}$;
    \item each pair of weights, $\mathit{weightLeft_i}$ and $\mathit{weightRight_i}$, resides in the regions $\mathit{w_i}$, $\mathit{1 \leq i \leq k}$;
    \item for each proximity sensor, $\mathit{prox_i}$, a compartment is defined, namely $s_i$, containing a single variable, $\mathit{x_{s_i,i}}$, $\mathit{1 \leq i \leq k}$;
    \item the products are calculated by two distinct programs, $\mathit{weightLeft_i \cdot prox_i}$, and $\mathit{weightRight_i \cdot prox_i}$, $\mathit{1 \leq i \leq k}$, in the compartments $\mathit{c_i}$.
\end{itemize}

The second model is an improvement on the first one. First we define the function

\[f(x) = 
\begin{cases}
   1, & \text{if } x = 0 \\
   0, & \text{otherwise}
\end{cases}
\]
\\

This function will be used in the equations describing the behavior of the model and in the production functions from the programs.\\

The equations describing the behavior are:
\begin{align*}
    \mathit{weightLeft} =& \displaystyle\sum_{i=1}^n \mathit{weightLeft_i} \cdot \mathit{prox_i} \\
    \mathit{weightRight} =& \displaystyle\sum_{i=1}^n \mathit{weightRight_i} \cdot \mathit{prox_i} \\
    \mathit{leftSpeed} =& \mathit{cruiseSpeed} \cdot \mathit{weightLeft + f(weightLeft)} \cdot \mathit{cruiseSpeed} \\
    \mathit{rightSpeed} =& \mathit{cruiseSpeed} \cdot \mathit{weightRight + f(weightRight)} \cdot \mathit{cruiseSpeed}
\end{align*}

The model is defined as follows:

$$\varPi \mathit{_{M_2} = (m, H, \mu, (Var_1, Pr_1, Var_1(0)),\dots,(Var_m, Pr_m, Var_m(0)))}$$

where:

\begin{itemize}
    \item $\mathit{m = 3k + 3, k = 6;}$
    \item $\mathit{H = \{s, w, s_c\} \cup \displaystyle\bigcup_{i=1}^{k} \{c_i, s_i, w_i\}}$;
    \item $\mathit{\mu = [[ [ []_{s_1} []_{w_1} ]_{c_1} \dots  [[]_{s_k} []_{w_k}]_{c_k}  []_{s_c} ]_{w} ]_{s};}$
    \item $\mathit{Var_{s} = \{x_{s_l}, x_{s_r}\}, Var_{w} = \{x_{w_l}, x_{w_r}, e_{w}\}, Var_{s_c} = \{ x_{s_c}\},}$ \\ \hspace*{-1mm} $\mathit{ Var_{c_i} = \{x_{c_i,s_l}, x_{c_i,s_r}, x_{c_i,w_l}, x_{c_i,w_r}, e_{c_i}\}, 1 \leq i \leq k,}$ \\ \hspace*{-1mm} $\mathit{ Var_{s_i} = \{x_{s_i,i}\}, 1 \leq i \leq k,}$ \\ \hspace*{-1mm} $\mathit{ Var_{w_i} = \{x_{w_i,w_l}, x_{w_i,w_r}, e_{w_i}\}, 1 \leq i \leq k}$;
    \item $\mathit{Var_i(0) = 0, 1 \le i \le k}$;
    \item $\mathit{Pr_{s} = \{0 \cdot x_{s_l} \cdot x_{s_r} \rightarrow 1|x_{s_l} + 1|x_{s_r}\}}$; \\
    $\mathit{Pr_{w} = \{x_{s_c} \cdot x_{w_l} + f(x_{w_l}) \cdot x_{s_c}|_{e_{w}} \rightarrow 1|x_{s_l},}$ \\ 
    \hspace*{12mm} $\mathit{x_{s_c} \cdot x_{w_r} + f(x_{w_r}) \cdot x_{s_c}|_{e_{w}} \rightarrow 1|x_{s_r}\}}$; \\
    $\mathit{Pr_{s_c} = \{x_{s_c} \rightarrow 1|x_{s_c}\}}$; \\
    $\mathit{ Pr_{c_i} = \{ x_{c_i,s_l} \cdot  x_{c_i,w_l}|_{e_{c_i}} \rightarrow 1|x_{s_l},}$\\ \hspace*{12mm} $\mathit{x_{c_i,s_r}  \cdot x_{c_i,w_r}|_{e_{c_i}} \rightarrow 1|x_{s_r} \}, 1 \leq i \leq k}$; \\
    $\mathit{Pr_{s_i} = \{3x_{s_i,i} \rightarrow 1|x_{s_i,i} + 1|x_{c_i,s_l} + 1|x_{c_i,s_r}\}, 1 \leq i \leq k}$; \\
    $\mathit{ Pr_{w_i} = \{ 2x_{w_i,w_l}|_{e_{w_i}} \rightarrow 1|x_{w_i,w_l} + 1|x_{c_i,w_l},}$\\ \hspace*{12mm} $\mathit{2x_{w_i,w_r}|_{e_{w_i}} \rightarrow 1|x_{w_i,w_r} + 1|x_{c_i,w_r} \}, 1 \leq i \leq k}$;
\end{itemize}

As we see from the definition of the $\varPi \mathit{_{M_1}}$ and its equations, in the proximity of an obstacle, roadside, the speed of the wheel on the side with the obstacle increases according to the weights. In certain circumstances, particularly related to the geometry of a tight road curve, the sum between travel speed and the weights results in a rotation in the opposite direction of the obstacle, but a slight forward motion still continues, causing the robot to leave the road. 

The second model, $\varPi \mathit{_{M_2}}$, overcomes this problem by calculating the product of the travel speed and the sum of the weights. In a similar situation, the second model performs an angular rotation, in the opposite direction to the obstacle, and when the proximity sensors no longer detect obstacles, it continues moving forward with a constant velocity. This behavior is modeled by compartment $\mathit{w}$.

In the next section, it can be seen that the $\varPi \mathit{_{M_2}}$ model has better behavior in similar situations compared to the first model.

\section{Simulation Results}

Having presented the way we integrate the tools along with the models we will detail in this section the testing stage. Ambiegen offers the possibility to configure the setup for its genetic algorithms, choosing the values for parameters like population size, number of generations, mutation rate or crossover rate. Also, aspects like the amount of time allocated for tests generation, map size, out of bound percent from which the test is considered as failed can be set easily from the command line, when executing the main Python file used in \cite{gambi2022sbst}. 

\begin{figure}[ht] 
  \begin{subfigure}[b]{0.5\linewidth}
    \centering
    \includegraphics[width=0.7\linewidth]{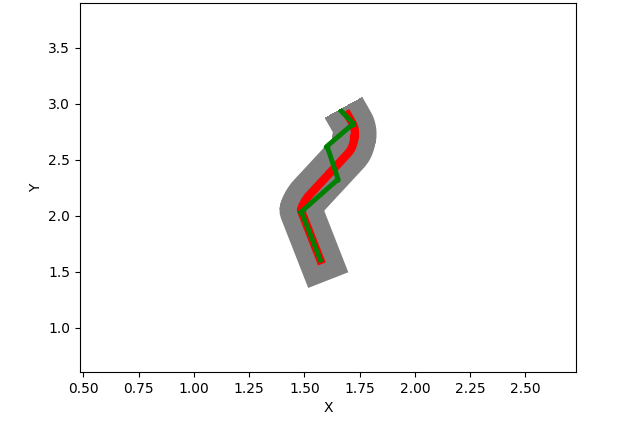} 
    \caption*{Test 1} 
    \label{fig7:a} 
    \vspace{1ex}
   \end{subfigure}
   \begin{subfigure}[b]{0.5\linewidth}
    \centering
    \includegraphics[width=0.7\linewidth]{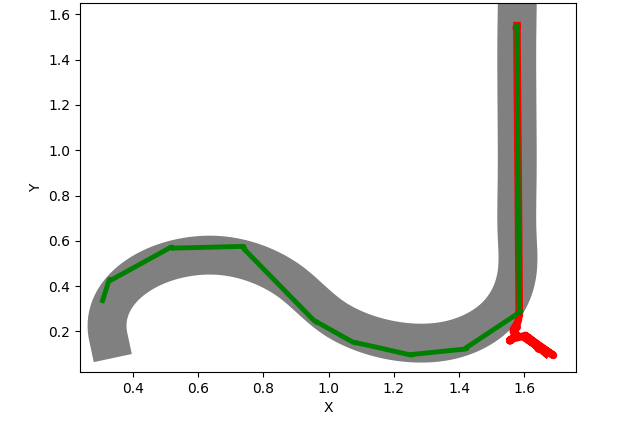} 
    \caption*{Test 2} 
    \label{fig7:b} 
    \vspace{1ex}
   \end{subfigure} 
  \begin{subfigure}[b]{0.5\linewidth}
    \centering
    \includegraphics[width=0.7\linewidth]{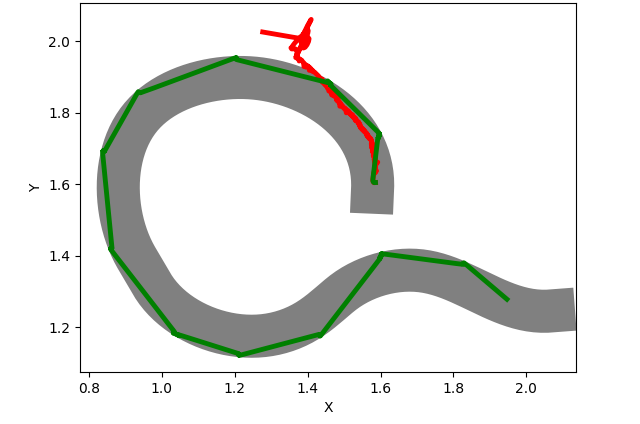} 
    \caption*{Test 3} 
    \label{fig7:c} 
  \end{subfigure}
  \begin{subfigure}[b]{0.5\linewidth}
    \centering
    \includegraphics[width=0.7\linewidth]{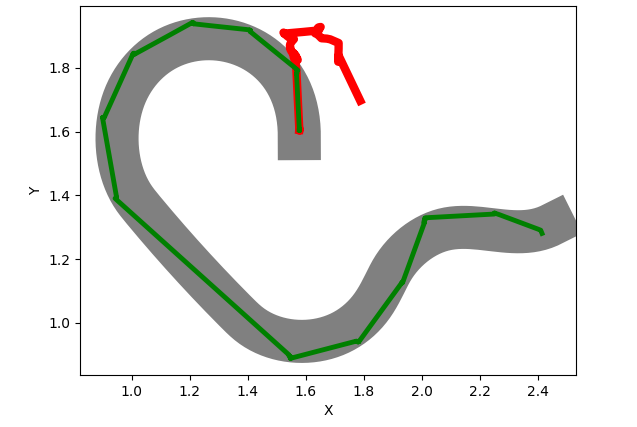} 
    \caption*{Test 4} 
    \label{fig7:d} 
  \end{subfigure} 
  \caption{Illustration of different road tests}
  \label{fig7} 
\end{figure}

After series of trials we kept the default values, so we used the population size 100 with 75 generations. Mutation rate is 0.4 and crossover rate is 1. These values can be changed from the internal configuration file of Ambiegen. From the command line we chose the time budget allocated to generation and execution to be 1800 seconds and the map size to 200x200 meters which is the default value. For the out of bound percentage we also kept the default value (95\%). After each simulation, Ambiegen exported roads spines coordinates as text files and we also plotted each generated road, as mentioned before. Then we could easily chose different roads based on the number of curves and their angle as the main criteria to diversify the tests that were supposed to be given to E-puck controller in Webots.

Next we report some roads along with the simulation results. The road is marked with grey, whilst the trajectory using the first model is represented with red. The trajectory of the second model (the improved one) is colored with green. 

In Figure 1 we can see the results of different roads tests. We observe that Test 1, being the simplest test of the above presented, is passed by both models. In the next scenarios, the complexity of the road increases and only the improved model can pass. 

From all the experiments, we noticed that the first model (the one marked with red in results) cannot pass a huge majority of tests with curves, whilst the improved model performs a rotation movement when the road is curved and for this reason it manages to advance until the end of the road. Additional tests were added to \cite{github}.
\section{Conclusions and Future Work}

In this paper we presented an approach to test different enzymatic numerical P systems models using modern tools and search-based generated tests. We evaluated our approach on a  research and educational robot called E-puck, virtually represented in Webots simulator. We set up our working environment incorporating the tools with different scripts created to ensure a smooth integration between them and also a better data processing. We formally described each model involved and then showed the differences between the lane-keeping controllers resulted when using each of them. 
As future work, we will investigate the possibility to dynamically calculate the values for weights, which are at the moment empirically assigned. Based on the controller behavior during the previous tests, the weights values will be automatically adapted. Also, we will try to develop a method to generate more
complex roads in order to better challenge the controllers.

\bibliographystyle{eptcs}
\bibliography{bib}
\end{document}